\documentclass[journal=ancac3,manuscript=article]{achemso}

\usepackage[version=3]{mhchem} % Formula subscripts using \ce{}
\usepackage{graphicx}
\usepackage{color}
\usepackage{epstopdf}
\author{Patrizia Borghetti}
\affiliation{Centro de F\'{\i}sica de Materiales CSIC/UPV-EHU-Materials Physics Center, Manuel Lardizabal 5, 20018-San Sebastian, Spain}
\alsoaffiliation{Donostia International Physics Centre, Manuel Lardizabal 4, 20018-San Sebastian, Spain}
\author{Ane Sarasola}
\affiliation{Departamento F\'{\i}sica Aplicada I, Universidad del Pa\'{\i}s Vasco, 20018-San Sebastian, Spain}
\author{Nestor Merino-Diez}
\affiliation{Centro de F\'{\i}sica de Materiales CSIC/UPV-EHU-Materials Physics Center, Manuel Lardizabal 5, 20018-San Sebastian, Spain}
\author{Guillaume Vasseur}
\affiliation{Donostia International Physics Centre, Manuel Lardizabal 4, 20018-San Sebastian, Spain}
\author{Luca Floreano}
\affiliation{CNR-IOM, Laboratorio TASC, Basovizza SS-14 Km. 163.5, 34149 Trieste, Italy}
\author{Jorge Lobo-Checa}
\affiliation{Instituto de Ciencia de Materiales de Arag\'on (ICMA), CSIC-Universidad de Zaragoza, E-50009 Zaragoza, Spain}
\alsoaffiliation {Departamento de F\'{\i}sica de la Materia Condensada, Universidad de Zaragoza, E-50009 Zaragoza, Spain}
\author{Andr\'es Arnau}
\affiliation{Centro de F\'{\i}sica de Materiales CSIC/UPV-EHU-Materials Physics Center, Manuel Lardizabal 5, 20018-San Sebastian, Spain}
\alsoaffiliation{Departamento F\'{\i}sica de Materiales, Universidad del Pa\'{\i}s Vasco, 20018-San Sebastian, Spain}
\alsoaffiliation{Donostia International Physics Centre, Manuel Lardizabal 4, 20018-San Sebastian, Spain}
\author{Dimas G. de Oteyza}
\affiliation{Donostia International Physics Centre, Manuel Lardizabal 4, 20018-San Sebastian, Spain}
\alsoaffiliation{Ikerbasque, Basque Foundation for Science, 48011 Bilbao, Spain}
\author{J. Enrique Ortega}
\affiliation{Centro de F\'{\i}sica de Materiales CSIC/UPV-EHU-Materials Physics Center, Manuel Lardizabal 5, 20018-San Sebastian, Spain}
\alsoaffiliation{Departamento F\'{\i}sica Aplicada I, Universidad del Pa\'{\i}s Vasco, 20018-San Sebastian, Spain}
\alsoaffiliation{Donostia International Physics Centre, Manuel Lardizabal 4, 20018-San Sebastian, Spain}
\email{enrique.ortega@ehu.es}

\title[An \textsf{achemso} demo]
{Symmetry, Shape and Energy Variations in Frontier Molecular Orbitals at Organic/Metal Interfaces: the Case of F$_4$TCNQ}
\begin{document}

\begin{tocentry}
\begin{center}
\includegraphics[width=35mm,angle=270,clip]{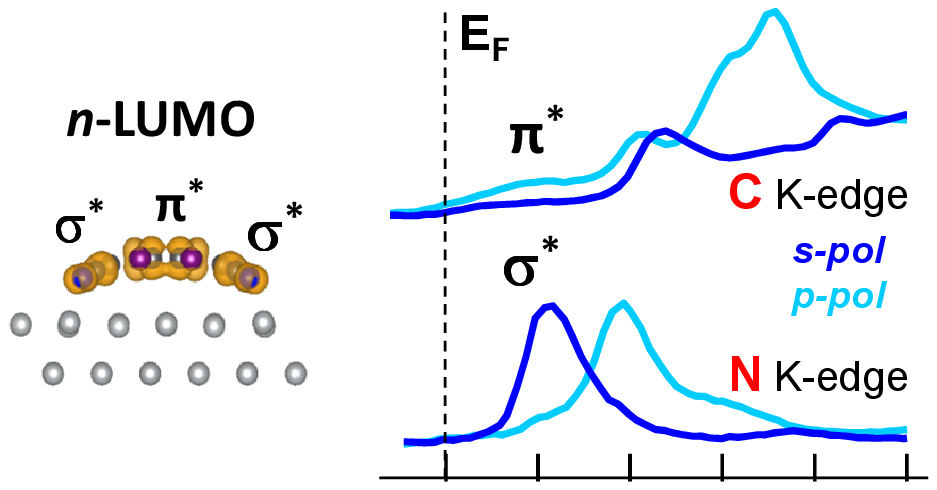}
\end{center}
\end{tocentry}


\pagebreak

\begin{abstract}
Near Edge X-ray Absorption, Valence and Core-level Photoemission and Density Functional Theory calculations are used to study molecular levels of tetracyano-2,3,5,6-tetrafluoroquinodimethane (F$_4$TCNQ) deposited on Ag(111) and BiAg$_2$/Ag(111). The high electron affinity of F$_4$TCNQ triggers a large static charge transfer from the substrate, and, more interestingly, hybridization with the substrate leads to a radical change of symmetry, shape and energy of frontier molecular orbitals. The Lowest Unoccupied Molecular Orbital (LUMO) shifts below the Fermi energy, becoming the new Highest Occupied Molecular Orbital (\emph{n}-HOMO), whereas the \emph{n}-LUMO is defined by a hybrid band with mixed $\pi^*$ and $\sigma^*$ symmetries, localized at quinone rings and cyano groups, respectively. The presence of Bi influences the way the molecule contacts the substrate with the cyano group. The molecule/surface distance is closer and the bond more extended over substrate atoms in F$_4$TCNQ/Ag(111), whereas in F$_4$TCNQ/BiAg$_2$/Ag(111) the distance is larger and the contact more localized on top of Bi. This does not significantly alter molecular levels, but it causes the respective absence or presence of optical excitations in F$_4$TCNQ core-level spectra.

\end{abstract}


\section{Introduction}

Organic/metal interfaces have become essential in all emerging electronic applications. Finding suitable materials combinations that fit the requirements of the desired device is a great challenge. One generally seeks sharp interfaces, in which substrate and organic layers retain their respective electronic properties. This generally defines a weak molecule-surface interaction scenario \cite{Braun09,Ishii99} characterized by low charge transfers, namely few tenths of electron per molecule, and interface electronics governed by the rigid alignment of molecular levels with respect to the metal Fermi level \cite{Goiri16,Hwang09,Heimel11,Oehzelt14,Savu15}. However, the design of new functional devices requires transferring substrate properties to the organic layer, such as magnetism, which in turn demands a stronger substrate-molecule interaction. However, enhanced molecule/surface interactions lead to larger hybridization effects \cite{Gonzalez08,Haeming12},and even induce structural disruptions of the substrate, such as atomic segregation \cite{Faraggi12}, making interface electronic states less predictable. Molecules with strongly donor or acceptor character may induce a sizeable interface charge transfer, of the order of one electron per molecule, which can both trigger profound electronic and conformational molecular changes \cite{Tseng10,Hauschild05,Heimel13} and modify the substrate properties \cite{Feng12}. These are attractive cases, which could prompt to the emergence of exotic interface properties, such as superconductivity or magnetism \cite{Kirtley08,Garnica13,Hsu13}.

In the quest for functional organic/metal interfaces, the combination of planar, aromatic molecules and noble metal surfaces has several advantages, particularly as model research systems. They generally form structurally sharp interfaces made of well-ordered supramolecular monolayers that lie flat on the metal surface. This facilitates the experimental analysis with surface sensitive microscopies and spectroscopies \cite{Tautz07,Dimas09,Ziroff10}, as well as theoretical modeling \cite{Hoffman13,Romaner09,Huang14}. Although most of such systems may still be considered weakly chemisorbed, one can find different degrees of interaction with the substrate \cite{Braun09}. As a prototype of strong acceptor molecules, tetracyanoquinodimethane (TCNQ) and his fluorinated counterpart tetracyano-2,3,5,6-tetrafluoroquinodimethane (F$_4$TCNQ, Fig. 1) have been largely investigated both at the interface with coinage metals \cite{Tseng10,Rangger09,Chen16,Faraggi12} or in combination with other organic materials \cite{Duhm07,Kirtley08}.

In this work, we examine in detail the large charge-transfer case of F$_4$TCNQ adsorbed on two chemically-different substrates, namely Ag(111) and BiAg$_2$/Ag(111). We combine Scanning Tunneling Microcopy (STM), X-ray Photoemission (XPS), Angle-resolved Photoemission (ARPES), Near-Edge X-ray Absorption (NEXAFS) and Density Functional Theory (DFT) calculations to assess differences in the interface structure, nature of the bonding and electronic properties. The observed transfer of 1.5 e$^-$ per molecule converts the primitive Lowest Unoccupied Molecular Orbital (LUMO) into the new Highest Occupied Molecular Orbital (\emph{n}-HOMO), but also a significant change in the wavefunction character of the new frontier molecular levels is observed. The \emph{n}-HOMO appears hybridized with substrate $s,p$ bands around the Fermi energy, and more interestingly, the upper edge of the gap (\emph{n}-LUMO band) exhibits a spatially modulated, mixture of $\sigma^*$ and $\pi^*$ symmetry orbitals. Differences among the two interfaces appear at the intimate cyano/metal contact, which has more localized character in F$_4$TCNQ/BiAg$_2$/Ag(111), and a higher degree of metallic delocalization in the F$_4$TCNQ/Ag(111) case.

\section{Results}

Due to its large electron affinity (-5.2 eV), the p-type organic dopant F$_4$TCNQ is widely used in charge-transfer multilayers and organic blends with potential opto-electronic applications \cite{Walzer07,Alves08}. The planar structure of the molecule is depicted in Fig. 1. Its functionality as an acceptor is determined by the four peripheral cyano groups and the four fluorine side atoms, all of which draw a significant amount of charge from the quinone ring. As it occurs in its parent TCNQ molecule \cite{Tseng10}, this quenches the aromatic character of the central hexagon, making the stand-alone molecule structurally rigid. The calculated projected density of states (PDOS) and the corresponding spatial charge distribution in frontier orbitals is also shown in Fig. 1.  The HOMO-1, HOMO, LUMO, LUMO+1, and LUMO+2 molecular orbitals have a marked $\pi$ nature, reflected in their nodal plane, whereas the LUMO+3 is the first unoccupied orbital that shows an in-plane local density of states (LDOS), namely a $\sigma$ character.

The adsorption of F$_4$TCNQ on Ag(111) leads to a number of changes in the molecular properties, defining radically new frontier \emph{n}-HOMO and \emph{n}-LUMO levels. This is immediately reflected in the DFT-calculated PDOS and spatial charge distribution, respectively shown in Figs. 2 (a) and (b). The large charge transfer from substrate to molecule leads to the shift across the Fermi energy of the empty LUMO peak of Fig. 1, which becomes the \emph{n}-HOMO of the molecule/surface system. This is clearly proven in NEXAFS, where the characteristic LUMO peak of the free-standing F$_4$TCNQ molecule vanishes (Fig. S1 in Supplementary Information). The Bader analysis indicates that approximately 1.5 e- have flown from the Ag(111) substrate to each F$_4$TCNQ molecule. Therefore, the \emph{n}-HOMO state is not entirely filled and extends above the Fermi level, as revealed in Fig. 2 (a). As discussed for the TCNQ/Cu(100) system \cite{Tseng10}, 1.5 e- represents a significant molecular charging that causes the aromatization of the central quinone ring and makes the molecule structurally flexible. This allows the molecule to bend, interact and hybridize with the substrate through the nitrogen lone pairs. The hybridization is also reflected in the charge density calculated at the \emph{n}-HOMO energy shown in the bottom of Fig. 2 (b). The $\pi$-symmetry orbitals extend to the surface through their symmetry-equivalent $p_z$-like states, particularly at atoms contacting the cyano groups. In contrast, the charge contour at the \emph{n}-LUMO energy [Fig 2 (b), top] does not show any significant density at the metal surface, although orbital changes with respect to the pristine molecule are significant. As shown in the PDOS plotted in Fig. 2 (a), the primitive LUMO+1, LUMO+2 and LUMO+3 levels reshuffle and overlap, thereby defining a frontier \emph{n}-LUMO band with the edge at 2.2-2.4 eV, which is made of a mixture of $\sigma^*$ and $\pi^*$ orbitals. This is in fact a coherent symmetry mixing, which affects the four individual molecular levels that make up the \emph{n}-LUMO band (see Fig. S2 in the Supporting Information file). As a consequence, the orbital symmetry of the \emph{n}-LUMO edge of the gap spatially varies from dominant $\sigma^*$ symmetry at the cyano group terminations to $\pi^*$ symmetry at the central quinone ring, as shown in the corresponding charge-density plot of Fig. 2 (b).

The profound changes of orbital energies and wavefunction character revealed in the DFT calculations are all mirrored in the NEXAFS spectra acquired at the C and N \emph{K}-edges. These are shown in Figs. 2 (c) and (d) for F$_4$TCNQ/Ag(111) and F$_4$TCNQ/BiAg$_2$ /Ag(111) interfaces, respectively. Due to the local atomic sensitivity of NEXAFS, peaks in the spectra correspond to the sequence of empty molecular levels probed at C and N atoms. Note the absolute electron energy scale used, which is determined from the respective C 1\emph{s} and N 1\emph{s} core-level binding energies (see core-levels in Fig. 4 below and Figs. S1 and S7 in the Supporting Information file), allowing us to make a direct comparison of the spectral weight at C and N positions for the same molecular orbital. A metal-like Fermi-edge is observed in all spectra close to zero energy, pointing at a significant metallization of the molecule \cite{Stoehr92}. The fact that a Fermi edge appears at zero energy justifies the use of the common electron energy scale, also indicating that core-hole exciton effects that affect the empty electronic levels in NEXAFS are minor \cite{Stoehr92,Johnson16}. Additionally, given the almost planar geometry of the organic layer, \emph{s} and \emph{p} polarization experiments test the respective $\sigma^*$ and $\pi^*$ orbital composition \cite{Stoehr92}. The dominant symmetry is clearly reversed for the frontier \emph{n}-LUMO peak at $\sim$ 1.1 eV when going from C to N edges, as qualitatively predicted by the DFT calculation shown in Fig. 2 (a). In contrast, the Fermi edge exhibits larger intensity in \emph{p}-polarization at both N and C edges. This confirms that the $\pi$ symmetry \emph{n}-HOMO level found at -0.97 eV in ARPES spectra (see Fig. 3) interacts with the metal $s,p$ band, broadening and tailing above the Fermi level. Note that except for a small energy shift of the \emph{n}-LUMO (at $\sim$ 1.4 eV in F$_4$TCNQ/BiAg$_2$/Ag(111)), the NEXAFS spectra for both F$_4$TCNQ/Ag(111) and F$_4$TCNQ/BiAg$_2$/Ag(111) interfaces are qualitatively equivalent.

By using ARPES, we next probe the occupied part of the energy level spectrum. Valence band spectra at F$_4$TCNQ/Ag(111) and F$_4$TCNQ/BiAg$_2$/Ag(111) interfaces are shown in Fig. 3 (a). When the interface with F$_4$TCNQ is formed, Shockley and Rashba-split bands that respectively characterize Ag(111) and BiAg$_2$/Ag(111) surfaces disappear (see Fig. S2 in Supplementary Information), giving place to non-dispersive molecular levels [Fig. 3 (a)] on a metallic background (Fermi edge). As found in other systems \cite{Temirov06,Tsirkin15}, charge transfer and hybridization usually pushes the Shockley surface state above the Fermi energy, which could be the case in F$_4$TCNQ/Ag(111). The absence of any surface band feature in F$_4$TCNQ/BiAg$_2$/Ag(111) contrasts with other molecule/BiAg$_2$/Ag(111) interfaces, where Rashba-split surface states survive under the organic molecular layer \cite{Walter14}. A comparison with DFT results in Fig. 2 (a) allows the direct assignment of peaks to \emph{n}-HOMO, \emph{n}-HOMO-1, and \emph{n}-HOMO-2 shown in Fig. 3 (b). From the peak positions in Fig. 3 (a) and the NEXAFS data of Figs. 2 (c) and (d), we elaborate the energy level diagram of Fig. 3 (b), which uses the Fermi level as a common zero energy reference. We have also added the vacuum level values for clean and molecule-covered substrates, determined from the low-energy cutoff in photoemission spectra, as well as the N 1\emph{s} peak energy, measured in core-level photoemission (see Fig. 4 below). We obtain a very similar sequence of molecular levels at both interfaces, with a quasi-rigid $\sim$0.2-0.3 eV shift to lower binding energies when going from F$_4$TCNQ/Ag(111) to F$_4$TCNQ/BiAg$_2$/Ag(111). Notably, the vacuum level is shifted by a similar amount (0.23 eV), suggesting that a similar molecule/surface interaction scenario holds in both cases, although a larger interface dipole exists at F$_4$TCNQ/BiAg$_2$/Ag(111). For occupied levels, the observed energy level spacing compares well with the DFT results in Fig. 2, whereas the interface \emph{n}-HOMO-\emph{n}-LUMO gap [2.07 eV in F$_4$TCNQ/Ag(111) and 2.02 eV for F$_4$TCNQ/BiAg$_2$/Ag(111)] agrees with the energy of the shake-up loss observed in core level peaks (see Fig. 4 and
the following discussion).The latter again corroborates the appropriate use of the electron energy scale in NEXAFS, as well as the effective metallic character of the organic layer in both cases, which leads to a large intramolecular screening of hole excitations \cite{Johnson16}.

Despite the strong resemblance of energies and symmetries of molecular levels in both F$_4$TCNQ/Ag(111) and F$_4$TCNQ/BiAg$_2$/Ag(111), important differences arise related to the way that the F$_4$TCNQ molecule contacts the respective metal surface. Since the amount of transferred charge is almost the same for both systems, the difference of $\sim$ 0.3 eV in the measured interface dipole can be assigned to a larger value of the molecule/surface distance, due to the longer N-Bi bond compared to the N-Ag bond (see Supporting Information). The DFT calculations shown in Fig. 4 (a) reveal notable differences in the degree of localization of the bond at the metal contact. In the top panel we represent the charge-density difference between the non-interacting F$_4$TCNQ+metal systems and the actual F$_4$TCNQ/metal interfaces, in a top-view perspective (see Supplementary Information for more details). Red and blue correspond to charge depletion and accumulation, respectively. The charge-density difference contour calculated for F4TCNQ/Ag(111) is in good agreement with that previously calculated by Rangger \emph{et al.} in \cite{Rangger09}. The comparison between the contour plots of F$_4$TCNQ/Ag(111) and F$_4$TCNQ/BiAg$_2$/Ag(111) in Fig. 4 (a) shows that differences in the charge distribution are minimal at quinone cores, but noteworthy around contact cyano groups. In particular, a more extended distribution of the charge-depleted area is observed at the metal substrate in the F$_4$TCNQ/Ag(111) case, in contrast to the more localized charge depletion effect found on top of the Bi atom at the F$_4$TCNQ/BiAg$_2$/Ag(111) interface (a side-view perspective is shown in Fig. S5 of the Supplementary Information). In agreement with the latter, the Bi 4\emph{f} core-level shows a $\sim$80 meV chemical shift towards high-binding energy [Fig. 4 (b)]. In the bottom panel of Fig. 4 (a) we zoom in the HOMO-LUMO gap region of the $\pi$ symmetry projected DOS of F$_4$TCNQ at both interfaces. In F$_4$TCNQ/Ag(111) we observe a flat and featureless background that fills up the \emph{n}-HOMO-LUMO gap, whereas the molecule gap in F$_4$TCNQ/BiAg$_2$ exhibits a discrete series of peaks, which is coherent (independently of the calculation parameters) with the sequence of $p_z$ levels found for Bi atoms (dotted lines). Such stronger and coherent modulation of the DOS at the interface is expected from a more localized character of the N-Bi interaction in F$_4$TCNQ/BiAg$_2$, in comparison to the delocalized bond formed between F$_4$TCNQ and the Ag(111) surface.

The distinct degree of localization of the N/metal bond may explain the striking differences in the N 1\emph{s} core-levels shown in Fig. 4 (c). Spectra correspond to the F$_4$TCNQ monolayer adsorbed on BiAg$_2$ (bottom), on Ag(111) (middle), and for a four layer thick film on Ag(111) (top). The latter shows the characteristic HOMO-LUMO ($\pi \rightarrow \pi^*$) shake-up satellite at $\sim$2.4 eV below the main peak \cite{Higo03,Tseng10}. At the BiAg$_2$ interface, the shake-up excitation is reduced to 1.80 eV and its intensity is notably enhanced, in contrast with the Ag(111) contact, where the shake-up is quenched. Similarly, a strong reduction of the satellite intensity is observed in the C 1\emph{s} spectrum of F$_4$TCNQ/Ag(111)(see Supporting Information). The intensity in shake-up satellites is proportional to the spatial overlap of all involved levels, namely core, HOMO and LUMO states \cite{Keane90,Keane91}, and inversely proportional to the energy of the excitation, namely the size of the HOMO-LUMO gap. Therefore, the large N 1\emph{s} shake-up intensity at the F$_4$TCNQ/BiAg$_2$/Ag(111) interface, compared to that of the F$_4$TCNQ multilayer, can be explained as due to a stronger localization of both \emph{n}-HOMO and \emph{n}-LUMO orbitals around N atoms in cyano groups, as well as to the effectively smaller \emph{n}-HOMO-\emph{n}-LUMO gap, as defined by the shake-up energy. On the other hand, the disappearance of the HOMO-LUMO shake-up loss in N 1$s$ at the F$_4$TCNQ/Ag(111) interface is surprising, since symmetry and energy of \emph{n}-HOMO and \emph{n}-LUMO molecular levels are similar for both interfaces (Fig. 2). The absence of the N 1\emph{s} shake-up satellite in F$_4$TCNQ/Ag(111) (also observed for TCNQ on Cu(100) \cite{Tseng10}) must therefore be related to the large spread and metallization of molecular orbitals at the cyano/substrate contact, which leads to the effective collapse of the optical gap. In reality, the calculations show metallic states inside the \emph{n}-HOMO-\emph{n}-LUMO gap at both interfaces [Fig. 4 (a)], but in the case of F$_4$TCNQ/BiAg$_2$/Ag(111) such gap states are the Bi-related $p_z$-resonances, which therefore appear not to affect optical excitations.

\section{Discussion}

F$_4$TCNQ /Ag(111) and F$_4$TCNQ/BiAg$_2$/Ag(111) are model abrupt
molecule/metal interfaces where a large 1.5 $e^-$/molecule charging occurs. NEXAFS, XPS, and ARPES provide us with the full spectrum of interface molecular levels and, in combination with a local molecule/metal DFT model, allows us understanding in detail the chemistry of the molecule/surface contact. Partial LUMO filling and metal hybridization occurs at other molecule/Ag(111) interfaces where the LUMO aligns close to $E_F$, such as PTCDA \cite{Schoell05,Zou06, Ziroff10, Duhm08, Romaner09} or CuPc \cite{Stadtmueller16,Borghetti14,Huang14}. For these systems, NEXAFS spectra show that the unoccupied molecular levels at low energies maintain a $\pi^*$  character both in the multilayer and monolayer form \cite{Zou06,Schoell05, Borghetti14}. By contrast, in the F$_4$TCNQ/metal system the high electron affinity of the molecule drives a much larger charging, which not only converts the LUMO of the free molecule into the new \emph{n}-HOMO \cite{Tseng10,Koch05}, but, as we show here, drastically alters the nature of all frontier levels with respect to the multilayer case \cite{Bassler00}. In particular, the new \emph{n}-LUMO appears defined by a combination of the pristine LUMO+1, LUMO+2 and LUMO+3 levels. As a consequence, we observe the spatial modulation of the orbital symmetry at the new \emph{n}-LUMO edge, varying from $\pi^*$ at the quinone center to a more dominant $\sigma^*$ symmetry at the cyano end group of the molecule.

Orbital symmetry at the molecule/metal interface determines the degree of molecule/surface coupling and, hence, influences hole and electron injection and transport. As discussed by Seideman \cite{Seideman16}, for planar molecule/metal contacts the relevant property is the symmetry with respect to the plane of the molecule for wave functions of both molecular levels and surface states that overlap in energy. $\sigma$ and $\sigma^*$ orbitals are symmetric with respect to the molecular plane, in contrast to the antisymmetric character of both $\pi$ and $\pi^*$ molecular states. For surface state wave functions that are symmetric with respect to the molecular plane coherent coupling is expected with $\sigma$ and $\sigma^*$ states, whereas antisymmetric surface states couple to $\pi$ and $\pi^*$ orbitals.  Coherent molecule/metal coupling at the interface is needed in order to texture molecular states with exotic substrate properties. This is claimed to be the case of C$_6$F$_6$ on Cu(111), where coherent coupling occurs between the molecule $\sigma^*$ orbital and the $s,p_z$-like image state, leading to molecular levels with metallic, nearly-free-electron character \cite{Dougherty12}. The present F$_4$TCNQ case is even more exotic, because  we have both $\pi^*$ and $\sigma^*$ symmetry at the same \emph{n}-LUMO energy, and hence coherent coupling is allowed with symmetric or antisymmetric substrate states. Such coupling scenario remains to be explored in the future.

The presence of Bi alters the molecular level spectrum of F$_4$TCNQ in a subtle way. Bi atoms intercalate the organic layer, acting as hole donors, that is, by varying Bi concentration at the interface the whole set of molecular levels shift to lower binding energy (see Supplementary Information file). Therefore, from the electronic transport point of view, a small change in hole and electron injection barriers can be smoothly tuned by adding Bi to the molecule/Ag(111) system. Notably, supramolecular ordering is not relevant, since different structural phases of F$_4$TCNQ exhibit identical molecular level spectra (see Supplementary Information file). Yet remarkable differences between Ag(111) and BiAg$_2$ interfaces are found at the cyano/metal contact, which affect optical excitations. In the pristine Ag(111) surface the cyano/metal distance is shorter, and the molecule/metal bond is largely spread, being both characteristic signatures of the strong metallization of the contact. This leads to the effective disappearance of the HOMO-LUMO gap in the cyano group, which in turn explains the absence of the shake-up satellite in the N 1$s$ photoemission spectrum. In the Bi-doped interface, the N contact with the out-protruding Bi atoms becomes more localized. Moreover, DFT maps indicate that charge flows mostly from Bi atoms and less from the surface Ag. Such local Bi/cyano contact explains chemical shifts in Bi core-levels, as well as the discrete series of Bi-like $p_z$ resonances that appear inside the HOMO-LUMO gap. However, the latter is effectively preserved, since the $n$-HOMO-$n$LUMO excitation that arises at the cyano group remains, as probed in the N 1$s$ spectrum.

\section{Methods}

\subsection{Density Functional Theory}

Calculations for gas-phase and adsorbate systems were carried out within density functional theory (DFT) as implemented in the computer code VASP\cite{Kresse93,Kresse96,Kresse96b}, using the Perdew, Burke and Ernzerhof generalized gradient approximation for exchange and correlation (GGA-PBE)\cite{PBE96} and optB88-vdW functional to describe the non local van der Waals interaction\cite{Klim10}. The Kohn-Sham wave functions are expanded in the plane wave basis with a kinetic energy cutoff of 400 eV for both adsorbate systems. Periodic supercells slabs are made of four 3$\times$3 Ag atomic layers for the Ag(111) surface, and of three Ag atomic layers plus the $\sqrt3 \times \sqrt3$ $R30^{\circ}$ topmost superstructure alloy of BiAg$_2$ for the BiAg$_2$/Ag(111) system. Monkhorst-Pack k-point sampling of 4$\times$6$\times$1 was employed to describe the first Brillouin zone. The optimization of the geometries include the relaxation of all the atoms of the adsorbate and the topmost layer of the substrate until the forces were smaller than 0.05 eV/\AA. VESTA software\cite{Momm11} has been employed for the post processing of the volumetric data required to map the LDOS and induced charge densities.

\subsection{Sample preparation}

The Ag(111) surface was prepared by standard sputtering and annealing cycles. The BiAg$_2$ alloy was obtained by evaporating one-third of a monolayer of Bi from a Knudsen cell with the sample kept at 300K, followed by a gentle annealing to T$\sim$550 K. Tetracyano-2,3,5,6-tetrafluoroquinodimethane (F$_4$TCNQ, from Sigma-Aldrich) was deposited from resistively heated Knudsen cells at T=350 K on the sample kept at T = 300 K, in the case of the monolayer deposition, and at T=200 K for multilayer deposition. Bi and molecular thicknesses were determined by a quartz crystal microbalance, and the calibration corroborated by STM, or by detailed analysis of the relative core-level peak intensities in synchrotron radiation measurements.The structural and electronic analysis of the molecular layer was performed in two different ultra-high vacuum (UHV) systems: at a home-laboratory with STM and angular resolved photoemission spectroscopy (ARPES), and at the ALOISA beamline of the ELETTRA synchrotron in Trieste, Italy, for high-resolution photoemission spectroscopy and near edge x-ray absorption fine structure (NEXAFS). The quality of BiAg$_2$ alloys and molecular films was further checked by low energy electron diffraction (LEED) at the home laboratory and by reflection high-energy electron diffraction (RHEED) at ELETTRA.

\subsection{Near Edge X-ray Absorption Fine Structure and X-ray Photoemission}

NEXAFS spectra were acquired by measuring the partial electron yield with a channeltron detector equipped with a front grid polarized at a negative bias of -230 and -380 V (for the C 1s and N 1s thresholds, respectively) in order to reject low energy secondary electrons. The spectra were measured with an energy resolution set to 80 (C 1s) and 100 meV (N 1s), and calibrated to the corresponding $1s\rightarrow \pi^*$ gas phase transitions of CO and N$_2$ at $h\nu$ = 287.4 and 401.10 eV, respectively, as described in Ref. \cite{Bavdek08}. Since the manipulator is coaxial to the photon beam, the change from linear s-polarization (s-pol, i.e. light polarization plane parallel to the sample surface) to p-polarization (p-pol, i.e. light polarization plane perpendicular to the sample surface) is obtained by simply rotating the sample around the beam axis while keeping a constant grazing angle of 6$^{\circ}$, i.e. without varying the illuminated area on the sample. The XPS data were collected by means of a hemispherical electron energy analyzer in normal emission while keeping the sample at grazing incidence ($\sim 4 ^{\circ}$). Spectra are measured in p-polarization at $h\nu$= 530 eV, which corresponds to an overall resolution of 200 meV. The binding energy of core-level spectra is carefully calibrated taking the peak of Ag 3$d_{5/2}$ at 368.1 eV as an absolute reference. The fitting of all XPS was done using a Shirley background and Voigt integral functions (See Supporting Information for details).

\subsection{Valence Band Photoemission and Scanning Tunneling Microscopy}

ARPES measurements were performed at T = 150 K using the He I line (21.2 eV)
from a monochromatized gas discharge lamp and a SPECS Phoibos 150 electron analyzer with energy and momentum resolutions of 40 meV and 0.1$^{\circ}$, respectively. In this case, the binding energy of valence band spectra is calibrated taking the Fermi level as an absolute reference. STM images were measured at room temperature in a commercial Omicron VT-STM operated with Nanonis control electronics in constant current mode. The analysis of the STM images was performed with the freeware WSxM from Nanotec \cite{Horcas07}.

\section{Supplementary Information}

Comparison between the electronic states of monolayers and multilayers (Figure S1); quenching of surface bands upon F$_4$TCNQ adsorption (Figure S2); influence of the supramolecular structure and Bi concentration on the valence band structure (Figure S3 and S4); charge-density and molecule/surface height variations upon interface formation (Figure S5); F 1\emph{s} (Figure S6) and C 1\emph{s} (Figure S7) photoemission spectra and results of the fit analysis of N 1\emph{s} (Table S1) and C 1\emph{s} spectra (Table S2).

\begin{acknowledgement}

We acknowledge financial support from the Spanish Ministry of Economy (Grants MAT2013-46593-C6-4-P, MAT2016-78293-C6-6-R and FIS2016-75862-P) and Basque Government (Grants IT-621-13 and IT-756-13).
P. B. acknowledges financial support from the European Union's
Horizon 2020 research and innovation programme under the
Marie Sklodowska-Curie grant agreement No. 658056.

\end{acknowledgement}

\section{Contribution}

P. B., G. V., N. M, J. L.-C, L. F., D. de O., and J. E. O performed experiments and analyzed data. A. S. and A. A. performed calculations. D. de O., A. A., and J. E. O. designed the research project. J. E. O. wrote the paper. All authors thoroughly reviewed the article.

\section{Competing financial interests}

The authors declare no competing financial interests.

%\bibliography{references_molecules}

\providecommand{\latin}[1]{#1}
\providecommand*\mcitethebibliography{\thebibliography}
\csname @ifundefined\endcsname{endmcitethebibliography}
  {\let\endmcitethebibliography\endthebibliography}{}

%\bibitem{Romaner2007} Romaner, L.; Heimel, G.; Bredas, J.-L.; Gerlach, A.; Schreiber, F.; Johnson, R. L.; Zegenhagen, J.; Duhm, S.; Koch, N.; Zojer, E. Impact of Bidirectional Charge Transfer and Molecular Distortions on the Electronic Structure of a Metal-Organic Interface. Phys. Rev. Lett. 99 (2007) 256801.

\pagebreak

\section{Figures}

\begin{figure}[htbp!]
\includegraphics[bb=12 100 430 800,angle=270, scale=0.8]{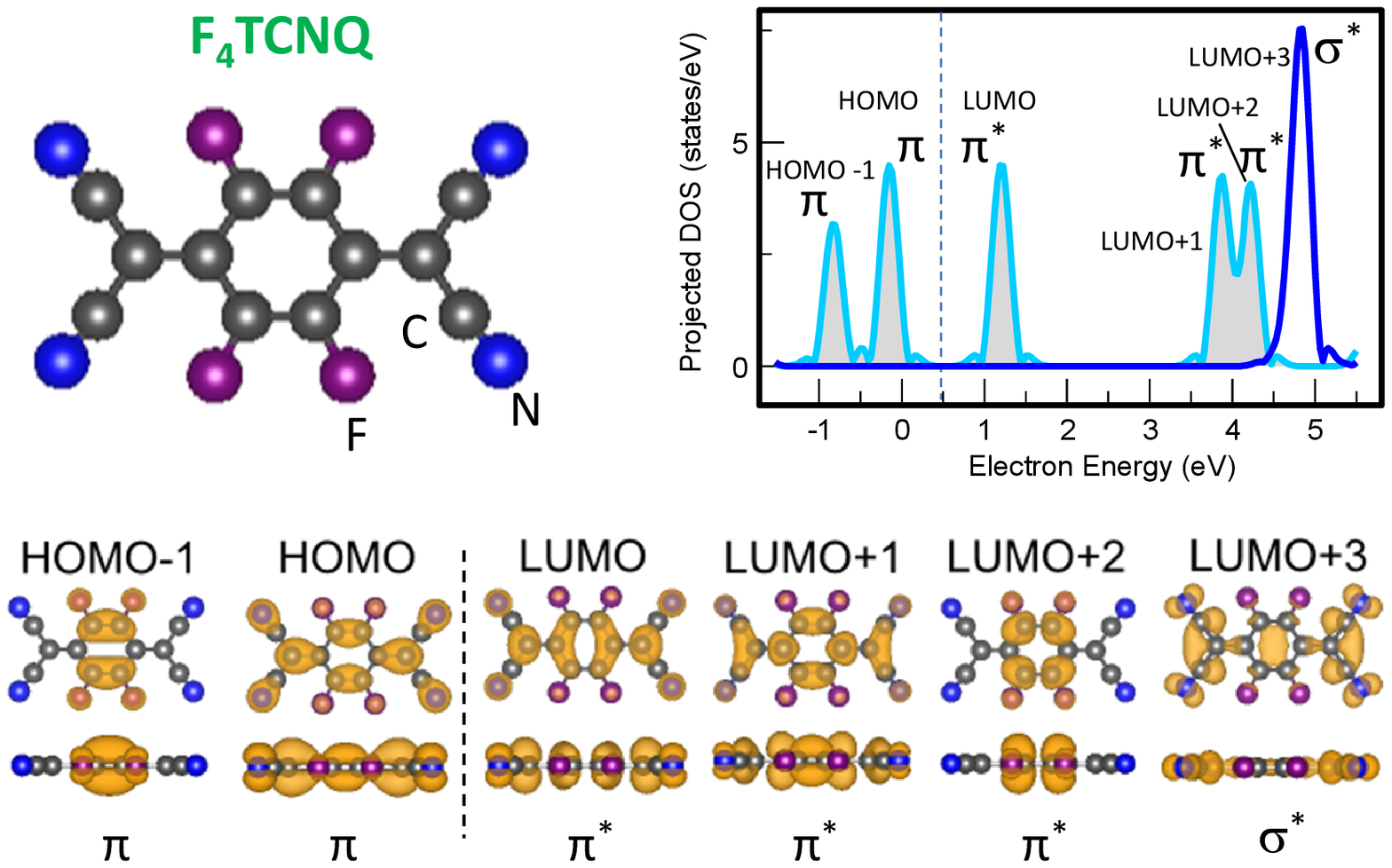}
\caption{\label{Fig1} \textbf{Structure and orbitals of free-standing F$_4$TCNQ}. Top-right, Density Functional calculation of the Projected Density of States for $\pi$ (light blue, shade) and $\sigma$ (dark blue) orbitals for the free-standing, organic acceptor F$_4$TCNQ (top-left). The front and side view of the corresponding charge density contours for the sequence of molecular levels is shown in the bottom.}
\end{figure}

\pagebreak

\begin{figure}[htbp!]
\includegraphics[bb=12 150 550 800,angle=270, scale=0.8]{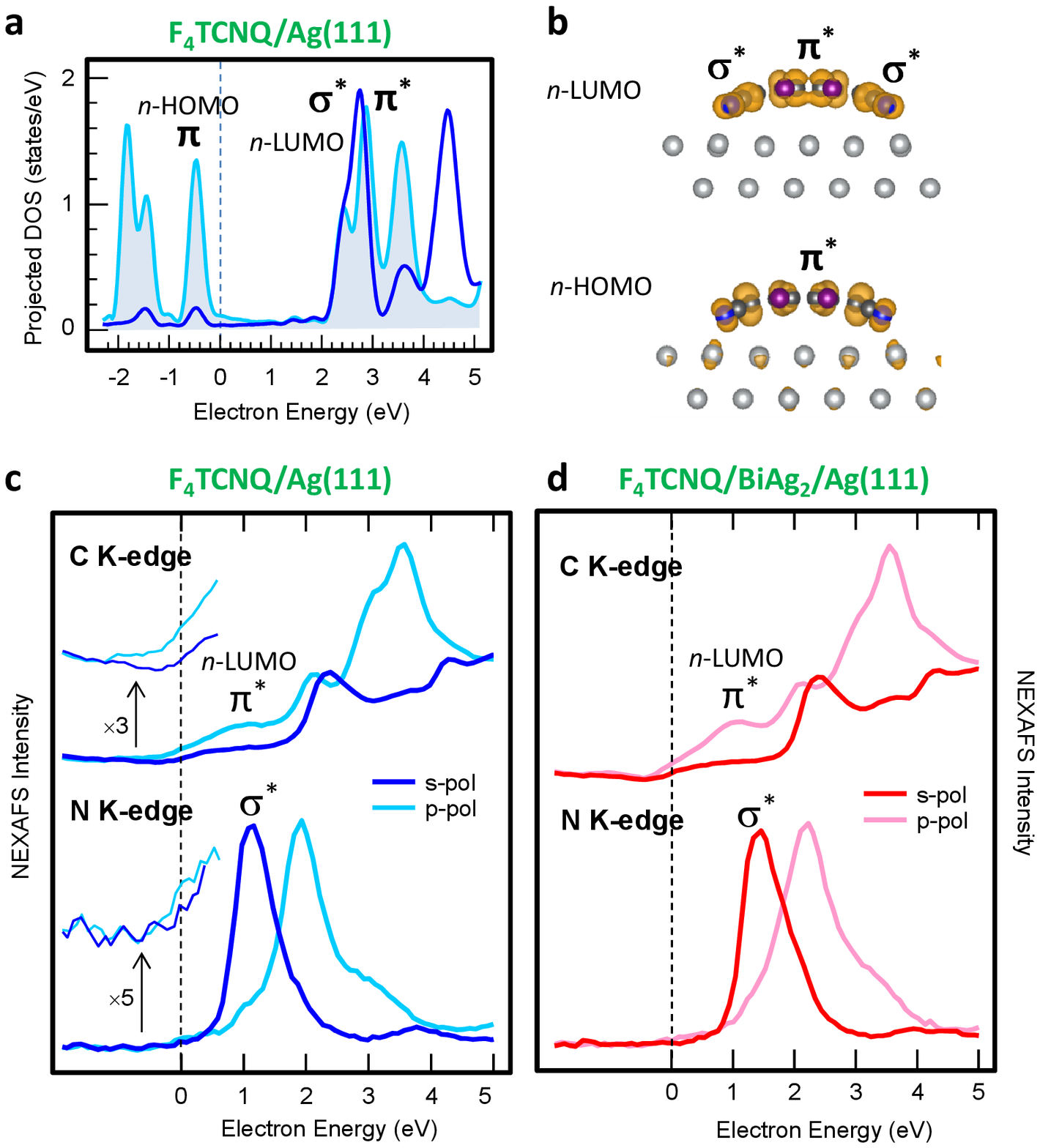}
\caption{\label{Fig2} \textbf{Unoccupied molecular levels and orbital character of adsorbed F$_4$TCNQ monolayers}. \textbf{(a)} Projected Density of States for $\pi$ (light blue, shade) and $\sigma$ (dark blue) orbitals of F$_4$TCNQ adsorbed on a Ag(111) surface. \textbf{(b)} Side view of the integrated charge density contours at n-LUMO and n-HOMO energies ($\Delta E$=1.2 eV). \textbf{(c-d)} NEXAFS spectra for C (top) and N (bottom) K-edges, taken with s-polarized (dark blue) and p-polarized (pale blue) light for F$_4$TCNQ monolayers adsorbed on Ag(111) and BiAg$_2$/Ag(111), respectively. The energy scale is determined from the binding energy of the corresponding XPS peaks (See Fig. 4 \textbf{(c)} and Supplementary Information). The p-polarized spectrum around zero energy is blown up in \textbf{(c)} to show the explicit Fermi edge onset.}
\end{figure}

\pagebreak

\begin{figure}[htbp!]
\includegraphics[bb=12 150 450 800,angle=270, scale=0.8]{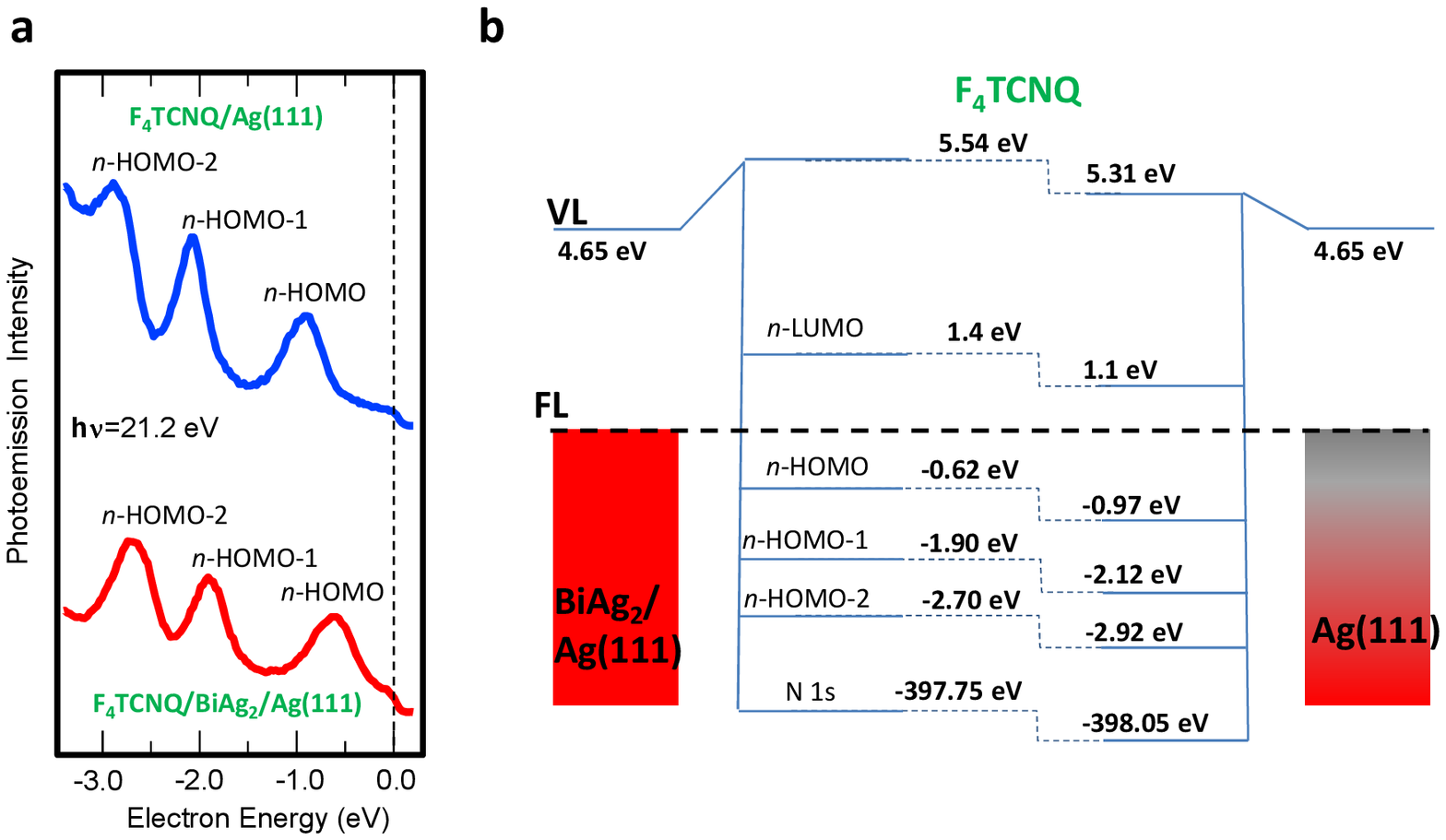}
\caption{\label{Fig3} \textbf{Molecular level spectra for adsorbed F$_4$TCNQ monolayers}. \textbf{(a)} Valence band Photoemission Spectra for F$_4$TCNQ monolayers adsorbed on Ag(111) (top) and BiAg$_2$/Ag(111) (bottom). Data correspond to Angle Resolved Photoemission Spectra (see Supplementary Information) integrated over $\pm 7^{\circ}$ around normal emission. The photon energy is set to 21 eV and the sample temperature to 150 K. \textbf{(b)} Chart of frontier molecular levels for F$_4$TCNQ monolayers on Ag(111) (right) and BiAg$_2$/Ag(111) (left), as determined from the NEXAFS (n-LUMO), ARPES (n-HOMO), and XPS (N 1s) data in Figs. 2 \textbf{(c-d)}, 3 \textbf{(a)}, and 4 \textbf{(c)}. The vacuum level position has been determined from the low energy cut-off shift measured in valence band photoemission, and using known work-function values of Ag(111) (4.65 eV) and BiAg$_2$/Ag(111) (4.65 eV) as reference.}
\end{figure}

\pagebreak

\begin{figure}[htbp!]
\includegraphics[bb=12 100 450 800,angle=270, scale=0.7]{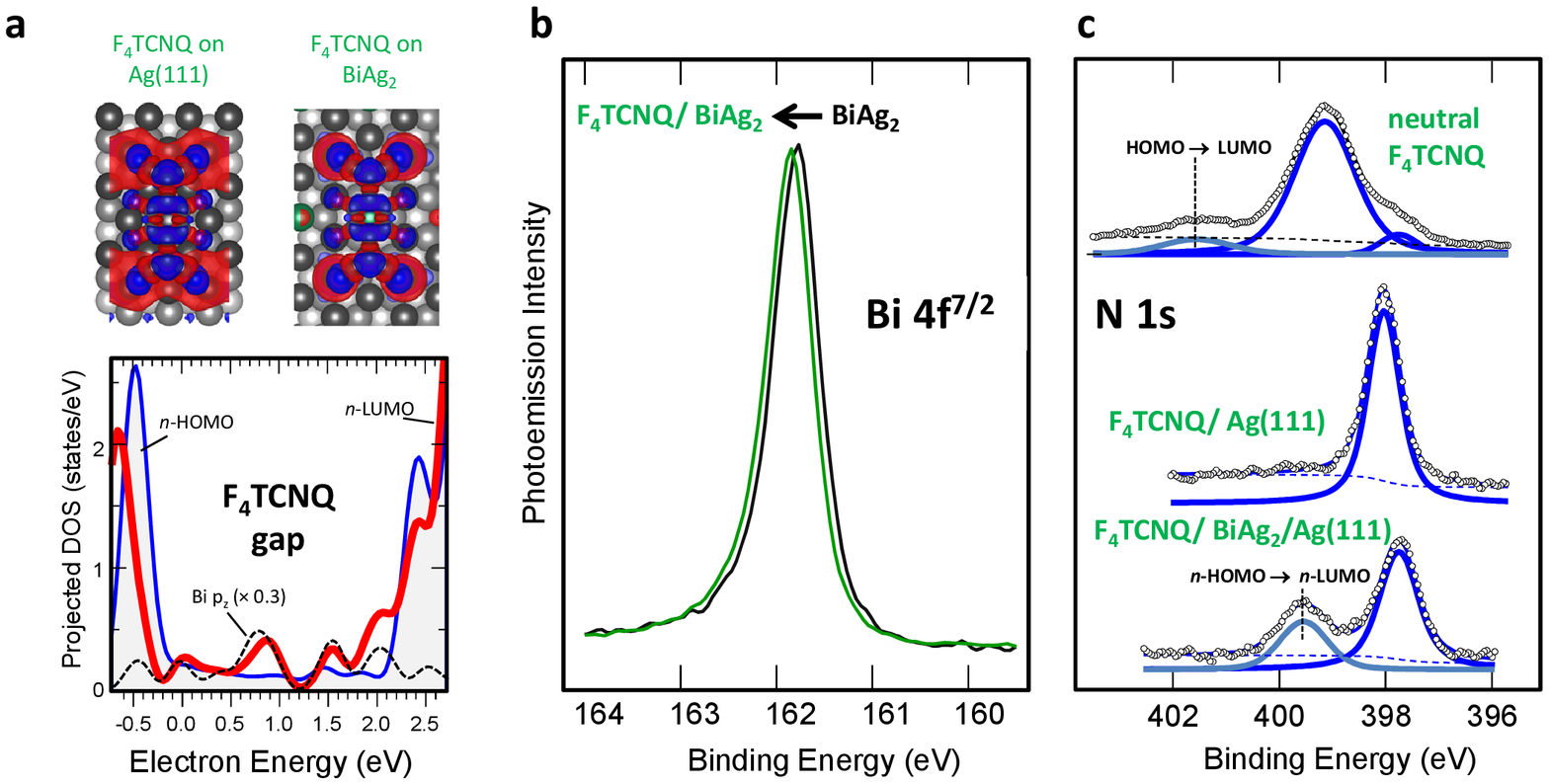}

\caption{\label{Fig4} \textbf{Analysis of the F$_4$TCNQ/metal contact for Ag(111) and BiAg$_2$/Ag(111)}. \textbf{(a)} Top, charge-density difference contour calculated for F$_4$TCNQ/Ag(111) (left) and F$_4$TCNQ/BiAg$_2$/Ag(111) (right) surfaces upon metal/contact formation. Blue and red stand for charge gain and loss, respectively. Bottom, $\pi$-like symmetry projected DOS in the HOMO-LUMO band gap of F$_4$TCNQ on Ag(111) (blue, shadow)) and BiAg$_2$ (red). The dotted line corresponds to the $p_z$-like projection at Bi atoms. (b) Bi 4$f_{7/2}$ core level shift upon F$_4$TCNQ monolayer covering. (c) N 1$s$ core level for a F$_4$TCNQ thick film on Ag(111) (top) and the F$_4$TCNQ monolayer on BiAg$_2$/Ag(111) (bottom) and Ag(111) (center). The $\pi \rightarrow \pi^*$ shake up of the neutral molecule in the thin film disappears in the charged molecule contacting Ag(111), but remains at the BiAg$_2$/Ag(111) interface.}
\end{figure}

\end{document}